\documentstyle[aps]{revtex}

\begin{document}

\title{On the dual equivalence of the Born-Infeld-Chern-Simons
model coupled to dynamical U(1) charged matter}

\author{D. Bazeia,$^1$ A. Ilha,$^2$ J. R. S. Nascimento,$^1$ R. F. Ribeiro$^1$
and C. Wotzasek$^2$}

\address{$^1$Departamento de F\' \i sica, Universidade Federal
da Para\'\i ba\\ 58051-970 Jo\~ao Pessoa, Para\'\i ba, Brasil\\
$^2$Instituto de F\'\i sica, Universidade Federal do Rio de Janeiro\\
21945-970, Rio de Janeiro, Brazil}

\date{\today}

\maketitle

\begin{abstract}
We study the equivalence between a nonlinear self-dual model (NSD) with
the Born-Infeld-Chern-Simons (BICS) theory using an iterative gauge embedding
procedure that produces the duality mapping, including
the case where the NSD model is minimally coupled to dynamical, U(1) charged
fermionic matter.
The duality mapping introduces a current-current interaction term while at
the same time the minimal coupling of the  original nonlinear self-dual
model is replaced by a non-minimal magnetic like coupling in the BICS side.
\end{abstract}
\bigskip

This work is devoted to the study of duality symmetry in the context of the
Born-Infeld-Chern-Simons (BICS) theory \cite{KT}, a generalization of the
self-dual and Maxwell-Chern-Simons (MCS) duality, firstly shown by Deser and
Jackiw \cite{DJ} long time ago using the formalism of master Lagrangian. To
this end we shall adopt a new gauge embedding formalism \cite{IW,AINRW} that
is alternative to the master Lagrangian approach.

Duality deals with the equivalence between models that describe the same
physical phenomenon.
This is a symmetry that is playing an important role in nowadays physics.
It has received a spate of interest in recent research in diverse areas in
field theory such as, supersymmetric gauge theories \cite{SW}, string
theories \cite{EK},
sine-Gordon model \cite{DC}, statistical systems \cite{JLC} and, in the
context of condensed matter models, applied for instance to planar high-T$_C$
materials, Josephson junction arrays \cite{DST} and Quantum Hall
Effect \cite{MS}.
The existence of such a
symmetry within a model may well have interesting consequences - it can
be used to derive (exact) non perturbative results since swapping opposite
coupling constant regimes allows a perturbative investigation of theories
with large coupling constants.
Moreover, it may be used to obtain
information on the corresponding phase diagram, as it has been done
in \cite{JLC} where the relevant duality symmetry has been related
to the modular group which, in that context, can be viewed as a
generalization of the old $Z_2$
Kramers-Wannier duality of the 2-dimensional Ising model.

Another possibility, which is closely related to our interest here,
is the odd-dimensional duality involving Chern-Simons term (CST) \cite{CST},
whose paradigm is the equivalence between Self-Dual and Maxwell-Chern-Simons
theories in (2+1) dimensions.
This is a very special situation since the presence of the topological and
gauge invariant Chern-Simons
term  is responsible for the essential features manifested by
the three dimensional field theories, such as parity breaking and anomalous
spin. It produces deep insights in unrelated areas in particle physics and
condensed matter both from the theoretical and phenomenological points of
view \cite{DJT}, as mentioned above.
The high temperature asymptotic of four dimensional
field theory models and the understanding of the universal behavior of the
Hall conductance in interacting electron systems also stand as important
illustrations of this topic.

Investigations of duality equivalence in D=3 involving CST has had a long
history. In 1984, Deser and Jackiw \cite{DJ} introduced the concept of master
Lagrangian to prove that there are two different ways to describe the
dynamics of a single, freely propagating spin one massive mode, either from
the SD \cite{TPvN} or from the MCS theories \cite{DJT}. In particular this
result has been of great significance in order to extend the bosonization
program from two to three dimensions with important phenomenological
consequences \cite{boson}.

On the other hand, the interest in the Born-Infeld theory \cite{BI} was recently
revived in the investigations related to fluid mechanics and D-branes.
For instance, both the relativistic Poincar\'e-invariant Born-Infeld model
and the non-relativistic Galileo-invariant Chaplygin gas in $(d,1)$
space-time dimensions descend from the very same reparametrization-invariant
Nambu-Goto d-brane action in $(d+1,1)$ dimensions \cite{jac99}. Also, in
string theory it is well known that the Dp-brane is governed by an action
$S_{Dp}$ which is the sum of the Born-Infeld action with the Chern-Simons
term, known as Born-Infeld-Chern-Simons (BICS) theory. For p=2 and 4,
$S_{Dp}$ can be derived from the actions of M2 and M5-brane in
M-theory \cite{dA}, and for p=0 it is obtained as the effective
action of the Kaluza-Klein mode of the gravity multiplet in the
eleven dimensional super-gravity. BICS theory is interesting in its
own right since it gives the MCS theory as a limiting case. In this
respect it becomes important to investigate the existence and to establish
the dual equivalence with a generalized model having the SD model as a
limiting case.

In this work we study the Chern-Simons duality for field theories including
self-interaction and consider in particular the nonpolinomial BI action.
To this end we apply an iterative gauge embedding procedure to construct
an invariant theory out of the nonlinear self-dual model leading to the
BICS, including the coupling to dynamical matter fields.
This is a gauge embedding procedure that changes the nature of the constraints
of a given noninvariant theory.  This is done with the inclusion of counter
terms in the action, built with powers of the Euler vector which is enough
to warrant the dynamical equivalence.
Such construction discloses, in the language of constraints, hidden
gauge symmetries in such systems.
One can then consider the non-invariant
model as the gauge fixed version of a gauge theory.
The former reverts to the latter under certain gauge fixing conditions.
By doing so we obtain a deeper and more
illuminating interpretation of these systems.
The associate gauge theory is therefore to be considered as the ``gauge
embedded" version of the original second-class theory. The advantage in
having a gauge theory lies in the fact that the underlying gauge symmetry
allows us to establish a chain of equivalence among different models by
choosing different gauge fixing conditions.

The question we pose now is: how to find, using the gauge embedding approach
for dual transformation, the (in principle, unknown) nonlinear SD theory that
is dual to the BICS model (a gauge theory)?  It should be clear that this is
the reversed of the usual situation described before where the noninvariant
model is known and one is looking for the (unknown) gauge embedded one. In
any event, to answer this question we start by reviewing the (direct) method
of gauge embedding \cite{IW,AINRW} for the traditional SD model,
\begin{equation}
\label{PB40}
{\cal L}_{SD}^{(0)} = \frac 12 m^2 f_\mu f^\mu - \frac m2 
\varepsilon ^{\mu\nu\rho} f_\mu\partial_\nu f_\rho \; ,
\end{equation}
where the superscript index is the iterative counter to be implemented here.
Our basic goal is to transform the hidden symmetry of the Lagrangian
(\ref{PB40}) into a local gauge symmetry
\begin{equation}
\label{PB47}
\delta f_\mu = \partial_\mu \epsilon
\end{equation}
with the lift of the global parameter $\epsilon$ into its local form, i.e.
\begin{equation}
\label{PB48}
\epsilon \to \epsilon(x,y,t)
\end{equation}
The method works by looking for an (weakly) equivalent description of this
theory which may be obtained by adding to the original Lagrangian
(\ref{PB40}) a function of the Euler vector $K_{\mu}$
\begin{equation}
\label{PB51}
{\cal L}_{SD}^{(0)} \to {\cal L}_{SD}^{(0)} + f(K_\mu)
\end{equation}
such that it vanishes on the space of solutions of (\ref{PB40}), i.e.,
$f(0)=0$. To find the specific form of this function that also induces
a gauge symmetry into ${\cal L}_{SD}^{(0)}$ we work out iteratively.
To this end we take the variation of ${\cal L}_{SD}^{(0)}$ to find the Euler
vector as
\begin{equation}
\label{PB50}
K^\mu = m^2 f^\mu - m \epsilon^{\mu\nu\lambda}\partial_\nu
f_\lambda
\end{equation}
whose kernel gives the equations of motion of the model.
Next we define a first-iterated Lagrangian as,
\begin{equation}
\label{PB60}
{\cal L}_{SD}^{(1)} = {\cal L}_{SD}^{(0)} - B_\mu K^\mu\, .
\end{equation}
where the Euler vector has been imposed as a constraint with $B_\mu$ acting
as a Lagrange multiplier.

The transformation properties of $B_\mu$ accompanying the basic field
transformation (\ref{PB47}) is chosen so as to cancel the variation of
${\cal L}_{SD}^{(0)}$, which gives
\begin{equation}
\label{PB70}
\delta B_\mu = \partial_\mu \epsilon
\end{equation}
A simple algebra then shows
\begin{eqnarray}
\label{PB80}
\delta {\cal L}_{SD}^{(1)} &=&  - B_\mu \delta K^\mu\nonumber\\
&=& - \delta \left(\frac {m^2}2 B^2\right) + m B_\mu \epsilon^{\mu\nu\lambda}
\partial_\nu \delta f_\lambda
\end{eqnarray}
where we have used (\ref{PB47}) and (\ref{PB70}).  Because of (\ref{PB47}), the
second term in the r.h.s. of (\ref{PB80}) vanishes identically leading to a
second iterated Lagrangian,
\begin{equation}
\label{PB90}
{\cal L}_{SD}^{(2)} = {\cal L}_{SD}^{(1)} + \frac {m^2}2 B^2
\end{equation}
that is gauge invariant under the combined local transformation of
$f_\mu$ and $B_\mu$.

We have therefore succeed in transforming the global SD theory into a locally
invariant gauge theory.  We may now take advantage of the Gaussian
character of the auxiliary field $B_\mu$ to rewrite (\ref{PB90})
as an effective action depending only on the original variable $f_\mu$.
To this end we use (\ref{PB90}) to solve for the field $B_\mu$ (call this
solution $\bar B_\mu$), and replace it back into (\ref{PB90}) to find
\begin{eqnarray}
\label{PB100}
{\cal L}_{eff}&=&{\cal L}_{SD}^{(2)}\mid_{B_\mu = \bar B_\mu} \;=
{\cal L}_{SD}^{(0)} - \frac 1{2m^2} K^2 \nonumber\\
&=& \frac 12 m^2 f_\mu f^\mu - \frac m2 
\varepsilon^{\mu\nu\rho} f_\mu\partial_\nu f_\rho\nonumber\\
& & -\frac 1{2m^2} \left[m^2 f^\mu - m \epsilon^{\mu\nu\lambda}
\partial_\nu f_\lambda\right]^2
\end{eqnarray}
which can be rewritten to give the MCS theory,
\begin{equation}
\label{BI40}
{\cal L}_{MCS} = -\frac 14  F_{\mu\nu} F^{\mu\nu} - \frac m2 
\varepsilon ^{\mu\nu\rho} A_\mu\partial_\nu A_\rho \; .
\end{equation}
It becomes clear from the above derivation that the difference between these
two models is given by a function of the Euler vector of the SD model that
vanishes over its space of solutions.  This establishes the
dynamical equivalence between the SD and the MCS theories.

Next, still in the direct situation, we consider the inclusion of
self-interactions. To be specific let us consider a SD model with a quartic
term,
\begin{equation}
\label{BI50}
{\cal L}_{2} = {\cal L}_{SD} + \frac {\lambda}4  \left(f_\mu f^\mu\right)^2\; .
\end{equation}
The kinematics of the theory is controlled by the CST with the dynamics being
determined by the potential that includes the mass term and the quartic
self-interaction.  Diagrammatically the SD model determines the basic
propagator while the quartic term gives the vertex with strength $\lambda$.
The basic idea of our approach is to reduce the self-interaction to a
trilinear term.  To this end we shall use the technique of auxiliary fields
to reduce the interaction term to the form, $\sigma f_\mu f^\mu$, that may
be treated as in the case above with the mass term being field dependent,
plus a function of the auxiliary field.  For the quartic term, only one
auxiliary field is necessary,
\begin{equation}
\label{BI60}
\frac {\lambda}4  \left(f_\mu f^\mu\right)^2\ \to \frac 12 {\sigma}
f_\mu f^\mu + \frac 1{\lambda} \sigma^2 .
\end{equation}
This operation has the effect of smooth out the quartic
vertex very much like the old Fermi theory with the introduction of an
intermediary field $\sigma$.

It is easy now to find the CS dual of this interacting SD model following
the general prescription developed above.  To this end we compute the Euler
current $K_\mu$ relative to gauge transformations of the basic field
$f_\mu$ keeping fixed the auxiliary field
\begin{equation}
\label{BI70}
K^\mu = \left(m^2+\sigma\right) f^\mu -
m \epsilon^{\mu\nu\lambda}\partial_\nu f_\lambda .
\end{equation}
The effective Lagrangian that results following the above procedure reads,
\begin{eqnarray}
\label{BI80}
{\cal L}_{eff} &=& {\cal L}_{SD}- \frac 1{2m^2+ \sigma} K_\mu^2 +
\frac 2{\lambda} \sigma^2 \nonumber\\
&=& m \epsilon^{\mu\nu\lambda}A_\mu\partial_\nu A_\lambda +
\frac 2{\lambda} \sigma^2 + \frac {F_{\mu\nu}F^{\mu\nu}}{2m^2+ \sigma}
\end{eqnarray}
The final step is to solve the model for the field $\sigma$.
The equation of motion gives
\begin{eqnarray}
\label{BI90}
\sigma\left(\sigma + m^2\right)^2 = F_{\mu\nu}F^{\mu\nu}
\end{eqnarray}
Let us call $\bar\sigma$ the solution of this equation,
\begin{eqnarray}
\label{BI100}
\bar\sigma = \bar\sigma \left( F_{\mu\nu}^2 \right)
\end{eqnarray}
Bringing this solution back in the effective Lagrangian (\ref{BI80}) gives
the $\sigma$(Maxwell)-Chern-Simons theory that is dual to the quartic
self-interacting SD theory. 
\begin{eqnarray}
\label{BI82}
{\cal L}_{\sigma{CS}} = m \epsilon^{\mu\nu\lambda}
A_\mu\partial_\nu A_\lambda+\frac 2{\lambda}\bar\sigma
\left( F_{\mu\nu}^2 \right) + \frac { F_{\mu\nu}^2}{2\, m^2+
\bar\sigma \left( F_{\mu\nu}^2 \right)}
\end{eqnarray}
Of course a closed expression for $\bar\sigma$ in terms of the field tensor
$F_{\mu\nu}$ is hard obtain in this case. There is however a nonlinear case
where a closed expression is possible. This will be considered
in the sequence.

We next consider the reversed situation mentioned above and discuss the
duality equivalence involving the BICS theory. Let us say that some (yet)
unknown nonlinear SD theory
\begin{equation}
\label{BI110}
{\cal L}_{NSD} = H\left( f_\mu f^\mu\right) - \frac 1{2m} 
\varepsilon ^{\mu\nu\rho} f_\mu\partial_\nu f_\rho \; ,
\end{equation}
is the dual to the BICS with $H\left( f_\mu f^\mu\right)$ being nonlinear
and possibly a nonpolinomial function of the $f_\mu f^\mu$.  Our strategy
is the assume that this term posses a linear representation, in terms of
$f_\mu f^\mu$, given by an auxiliary field representation as
\begin{equation}
\label{BI120}
{\cal L}_{NSD} = g\left(\sigma\right) + \frac 1{2}
\sigma f_\mu f^\mu - \frac 1{2m}\varepsilon ^{\mu\nu\rho}
f_\mu\partial_\nu f_\rho \; .
\end{equation}
The function $g\left(\sigma\right)$ is then determined such that the dual
of (\ref{BI110}) gives the BICS theory.  To this end we repeat the steps
above and compute the Euler vector,
\begin{equation}
\label{BI130}
K^\mu = {\sigma} f^\mu -
\frac 1m \epsilon^{\mu\nu\lambda}\partial_\nu f_{\lambda}
\end{equation}
which allows us to find an effective action as
\begin{eqnarray}
\label{BI140}
{\cal L}_{eff} &=& {\cal L}_{NSD}-\frac{1}{2\sigma}K_\mu^2 \nonumber\\
&=& g\left(\sigma\right) + \frac 1{2m}
\varepsilon ^{\mu\nu\rho} A_\mu\partial_\nu A_\rho -
\frac{1}{4m^2} \frac 1\sigma F_{\mu\nu}F^{\mu\nu} .
\end{eqnarray}
We have also relabeled $f_\mu \to A_\mu$ to reflect the inherent gauge invariance of the system.
Next we solve for the field $\sigma$ in terms of $F_{\mu\nu}^2 $ to get,
\begin{eqnarray}
\label{BI150}
\left[\sigma^2  \frac d{d\sigma}g (\sigma)\right]_{\sigma = \bar\sigma}+
\frac{1}{4m^2} F_{\mu\nu}^2 = 0
\end{eqnarray}
We must look at a conditional solution for this equation such that its
substitution back into (\ref{BI140}) should give the BI action,
\begin{eqnarray}
\label{BI160}
g\left(\bar\sigma\right) - \frac{1}{4m^2} \frac 1{\bar\sigma}
F_{\mu\nu}^2 = \beta^2 \sqrt{1 - \frac 1{2\, m^2\beta^2}F_{\mu\nu}^2 }
\end{eqnarray}
The solution for these equations is given by 
\begin{eqnarray}
\label{BI170}
g\left(\sigma\right) = a \,\sigma + b\, \frac 1\sigma
\end{eqnarray}
with the constants $a$ and $b$ being determined by (\ref{BI160})
self-consistently as,
\begin{eqnarray}
\label{BI180}
a=b=\frac {  \beta^2}{2} .
\end{eqnarray}
Bringing this solution back into (\ref{BI140}) gives 
\begin{eqnarray}
\label{BI140a}
{\cal L}_{BICS} = \beta^2\sqrt{1 - \frac 1{2\, m^2\beta^2}F_{\mu\nu}^2 }
+\frac 1{2m}
\varepsilon ^{\mu\nu\rho} A_\mu\partial_\nu A_\rho .
\end{eqnarray}
which is the BICS theory we are looking for.

We are now in position to find the nonpolinomial function $H(f_\mu^2)$
in (\ref{BI110}), that fixes the dual to the BICS theory, by bringing the
solution (\ref{BI150})-(\ref{BI180}) back into (\ref{BI120}) and solving the
equation of motion of the Lagrangian for the auxiliary field which, in this
case, has a closed exact solution as,
\begin{eqnarray}
\label{BI190}
\tilde\sigma =  \frac 1{\sqrt{1+
\frac{f_\mu f^\mu}{\beta^2}}}\; .
\end{eqnarray}
The auxiliary field is then eliminate from the theory by bringing it back
into the Lagrangian (\ref{BI140}) to produce,
\begin{equation}
\label{BI200}
{\cal L}_{NSD}=\beta^2{\sqrt{1+\frac{f_\mu f^\mu}{ \beta^2}}}-\frac 1{2m} 
\varepsilon ^{\mu\nu\rho} f_\mu\partial_\nu f_\rho \; .
\end{equation}
It is a simple algebra to show that indeed the equation of motion derived
from ${\cal L}_{NSD}$ and ${\cal L}_{BICS}$ coincide. This solution has
been found recently by Tripathy and Khare \cite{KT} by looking for a
generalization of the equations of motion of the SD and MCS actions
that have the correct limit for large values of $\beta$.

We are now in position to consider the minimal coupling of the NSD model
with dynamical fermions.  This generalizes the result of
Gomes et al \cite{GMdS} into the BI context.
Let us consider a Dirac field minimally coupled to a vector field specified
by the NSD model \cite{GMdS}, so that the Lagrangian becomes
\begin{equation}
\label{PBi40}
{\cal L}_{min}^{(0)} = {\cal L}_{NSD}  - e f_\mu J^\mu + 
{\cal L}_{D}  \; , 
\end{equation}
where $J^\mu = \bar{\psi}\gamma^\mu \psi$, $M$ is the fermion mass and the
superscript index is the iterative counter. Here the Dirac Lagrangian is,

\begin{equation}
\label{PB45}
{\cal L}_{D} =  \bar{\psi}(i\partial\!\!\! /  -M)\psi \; , 
\end{equation}
To implement the gauge embedding we follow the usual track and compute the
Euler vector for the Lagrangian ${\cal L}_{NSD}$ given by (\ref{BI120}),
now in the presence of the fermionic current,
\begin{equation}
\label{BI210}
K^\mu={\sigma}f^\mu-\frac 1m
\epsilon^{\mu\nu\lambda}\partial_\nu f_{\lambda} - e\, J^\mu
\end{equation}
The effective theory that comes out after the dualization procedure, with $f_\mu \to A_\mu$, is given
by (\ref{BI140}) with the Euler vector given by (\ref{BI210})
\begin{eqnarray}
\label{BI220}
{\cal L}_{eff} &=& g\left(\sigma\right) + \frac 1{2m}
\varepsilon ^{\mu\nu\rho} A_\mu\partial_\nu A_\rho \nonumber\\
&-& \frac 1\sigma\left[ \frac{1}{4m^2}  F_{\mu\nu}^2
+ \frac {e^2}{2} J_\mu^2 +\frac{e}m  \epsilon^{\mu\nu\lambda}\,
J_\mu\partial_\mu A_\lambda\right]
\end{eqnarray}
with $g(\sigma)$ given by (\ref{BI170}) and (\ref{BI180}).
Next the auxiliary field is eliminate in the usual way - from its
equation of motion we find
\begin{eqnarray}
\label{BI191}
\bar\sigma = \sqrt{1 - \frac 2{\beta^2} \left( \frac{1}{4m^2}  F_{\mu\nu}^2
+ \frac {e^2}{2} J_\mu^2 +\frac{e}m  \epsilon^{\mu\nu\lambda}\,
J_\mu\partial_\mu A_\lambda\right)}
\end{eqnarray}
which, when inserted back into (\ref{BI220}), produces the dual BICS theory
in the presence of dynamical fermions,
\begin{eqnarray}
\label{BI140b}
{\cal L}_{BICS}&=&\frac 1{2m} \varepsilon ^{\mu\nu\rho}
A_\mu\partial_\nu A_\rho \nonumber\\
& &+ 
{\beta^2}\sqrt{1 - \frac 2{\beta^2} \left( \frac{1}{4m^2}  F_{\mu\nu}^2
+ \frac {e^2}{2} J_\mu^2 +\frac{e}m  \epsilon^{\mu\nu\lambda}\,
J_\mu\partial_\mu A_\lambda\right)}
.
\end{eqnarray}
A simple inspection shows that the minimal coupling of the NSD model
was replaced by a nonminimal magnetic Pauli type coupling and also the
presence of a Thirring like current-current term.  These features are
characteristics of Chern-Simons dualities involving matter coupling
\cite{GMdS,AINRW}

\vspace{.5cm}

\noindent ACKNOWLEDGMENTS: This work is partially supported by CNPq, PRONEX, CAPES,
FAPERJ and FUJB, Brazilian Research Agencies.  CW thanks the Physics
Department of UFPB for the kind hospitality at the beginning of this
investigation.

\end{document}